\documentclass[twocolumn]{jpsj2} 

\title{Carrier-Induced Magnetic Circular Dichloism 
in the Magnetoresistive Pyrochlore Tl$_2$Mn$_2$O$_7$.}

\author{%
Hidekazu \textsc{Okamura}\thanks{E-mail: okamura@phys.sci.kobe-u.ac.jp}, 
Toshihisa \textsc{Koretsune}, 
Shin-ichi \textsc{Kimura}$^1$, 
Takao \textsc{Nanba}, \\
Hideto \textsc{Imai}$^2$, 
Yuichi \textsc{Shimakawa}$^2$\thanks{Present Address: 
Institute for Chemical Research, Kyoto University, Uji 611-0011.} and 
Yoshimi \textsc{Kubo}$^{2}$
}

\inst{%
Graduate School of Science and Technology, Kobe University, 
Kobe 657-8501 
\\
$^1$ UVSOR Facility, Institute for Molecular Science, 
Okazaki 444-8585
\\
$^2$ Fundamental Research Laboratories, NEC Corporation, 
Tsukuba 305-8501.
}

\recdate{\today}

\abst{%
Infrared magnetic circular dichloism (MCD), or equivalently 
magneto-optical Kerr effect, has been measured on the 
Tl$_2$Mn$_2$O$_7$ pyrochlore, which is well known for 
exhibiting a large magnetoresistance around the Curie 
temperature $T_C \sim$ 120~K.     A circularly polarized, 
infrared synchrotron radiation is used as the light source.   
A pronounced MCD signal is observed exactly at the plasma 
edge of the reflectivity near and below $T_C$.    
However, contrary to the conventional behavior of MCD for 
ferromagnets, the observed MCD of Tl$_2$Mn$_2$O$_7$ grows 
with the applied magnetic field, and not scaled with the 
internal magnetization.    
It is shown that these results can be basically understood in 
terms of a classical magnetoplasma resonance.     The absence 
of a magnetization-scaled MCD indicates a weak spin-orbit 
coupling of the carriers in Tl$_2$Mn$_2$O$_7$.     
We discuss the present results in terms of the microscopic 
electronic structures of Tl$_2$Mn$_2$O$_7$.     
}

\kword{
Tl$_2$Mn$_2$O$_7$, infrared magnetic circular dichloism, 
magnetoplasma, colossal magnetoresistance. 
}

\begin{document}
\maketitle

\section{Introduction} 
Physics of colossal magneto-resistance (CMR) phenomena 
has attracted much interest recently.  The most famous 
examples are probably the ferromagnetic perovskite 
manganites such as La$_{1-x}$Sr$_x$MnO$_3$.\cite{tokura}     
In these compounds, the so-called ''double exchange`` 
interaction and the strong Jahn-Teller effect due to 
Mn$^{3+}$ are responsible for the CMR and other 
interesting properties near the Curie temperature 
($T_C$).\cite{zener,millis}        
Another CMR compound that is equally interesting is the 
Tl$_2$Mn$_2$O$_7$ 
pyrochlore.\cite{shimakawa,subram,cheong,shimakawa2,shimakawa3,imai}     
This compound is also a ferromagnet, and its 
resistivity ($\rho$) drops rapidly upon cooling 
below $T_C$~$\sim$ 120~K.        
Near and above $T_C$, an external magnetic field 
($B$) of 7~T reduces $\rho$ by a factor of $\sim$ 10.     
Although these features appear similar to those 
for the perovskites, it has been shown that the 
underlying mechanism is very different: 
The spontaneous magnetization below $T_C$ is produced 
by the Mn$^{4+}$ sublattice through superexchange 
interaction, independently from the conduction electrons.   
The conduction band has large Tl 6$s$ 
and O~2$p$ components.\cite{shimakawa3,singh,mishra}      
The CMR mainly results from changes in the conduction band 
when the applied $B$ field induces a magnetization and 
hence a band polarization.     Such magneto-resistance 
mechanism has been considered\cite{shimakawa4,kataoka} 
to be analogous to that of europium chalcogenides 
such as EuO.

Our previous magneto-optical study of Tl$_2$Mn$_2$O$_7$ 
provided much 
information regarding its electronic structures.\cite{okamura}   
The optical reflectivity spectrum $R(\omega)$ at 295~K was 
similar to those for insulating oxides.   Upon cooling below $T_C$, 
however, $R(\omega)$ showed marked increases in the 
far-infrared region, with a clear plasma edge undergoing 
blue shifts with decreasing temperature ($T$).   
These changes clearly indicated a crossover to metallic 
electronic structures below $T_C$.  
$R(\omega)$ also showed large increases and plasma edge 
shifts when a strong magnetic field was applied near $T_C$, 
that were very similar to those caused by cooling.   
Based on the $R(\omega)$ data and the optical conductivity 
$\sigma(\omega)$ obtained 
from $R(\omega)$, we concluded that the CMR was caused 
by the appearance of a small conduction band, and that the 
$B$-induced changes in the electronic structure were very 
similar to those induced by cooling at $B$=0.   It was also 
found that the effective carrier density showed a universal 
scaling with $M^2$, where $M$ is the internal magnetization, 
over wide ranges of $T$ and $B$.

In this work we have studied the magnetic circular dichloism 
(MCD) of Tl$_2$Mn$_2$O$_7$ in order to obtain information about 
spin-dependent carrier dynamics below $T_C$.     
An MCD refers to a magnetically-induced difference between 
the optical responses 
to the incident light with right- and left-circular 
polarizations.          
In principle it can be a sensitive probe of spin-polarized 
electronic states in a magnetic material.        
Band calculations for ferromagnetic Tl$_2$Mn$_2$O$_7$ have 
predicted that the conduction electrons are completely 
spin-polarized.\cite{singh,mishra,shimakawa3}    
The carrier dynamics of such spin-polarized material is 
very interesting from both physical and technological 
points of view.    
A strong MCD has been observed exactly at the plasma 
edge of $R(\omega)$ in the infrared region, 
showing that the MCD arises from a small density of free 
carriers.  
The appearance of the MCD is closely related to the 
applied $B$, rather than $M$.     
We show that the observed MCD spectra can be basically 
understood based on a classical magneto-plasma model.   
The results are analyzed in terms of the 
electronic structures of Tl$_2$Mn$_2$O$_7$.

\section{Experimental}
We used the same Tl$_2$Mn$_2$O$_7$ sample, having 
$T_C \sim$~120~K, as that used in the previous 
report.\cite{okamura}   
The MCD experiments were done using a circularly 
polarized infrared synchrotron radiation as the 
light source, at the beam line BL6A1 of the UVSOR Facility, 
Institute for Molecular Science.\cite{kimura-1,kimura-2}    
Using synchrotron radiation source, one can obtain circularly 
polarized light over a wide range of photon 
energy,\cite{kimura-1,kimura-2,footnote-mcd} without 
additional optical elements such as a $\lambda$/4 plate.    
The reflectivity spectra of the sample were measured 
under a near-normal incidence, with magnetic fields 
applied perpendicular to the sample surface.     
An MCD under this condition is often referred to as the 
{\it magneto-optical Kerr effect}.\cite{MOKE}    
The incident light always had right circular polarization, 
and the magnetic field was applied either parallel 
($\vec{B}_+$) or antiparallel ($\vec{B}_-$) to the incident 
light.    In this work we define the MCD spectrum as 
\begin{equation}
\delta(\omega) = - \frac{R_+(\omega) - R_-(\omega)}{R_{avg}(\omega)}, 
\label{mcd}
\end{equation}
%
where $R_\pm(\omega)$ are the $R(\omega)$ spectra 
measured under $\vec{B}_\pm$ fields, respectively, and 
$R_{avg}(\omega)$ is the average of $R_\pm(\omega)$.  
Namely the MCD is the relative difference in $R(\omega)$ 
that results when the $\vec{B}$ direction is reversed 
for a fixed helicity of circularly polarized incident 
light.      
(This is equivalent to reversing the helicity of the 
circular polarization for a fixed $\vec{B}$ direction.)   
The MCD thus defined is 
directly related to the Kerr ellipticity, $\eta_K(\omega)$, 
as $\delta(\omega) = -4 \eta_K(\omega)$.\cite{MOKE}     
The Kerr rotation angle $\theta_K(\omega)$ is given as 
%
\begin{equation}
\theta_K(\omega) = - [\theta_+(\omega) - \theta_-(\omega)]/2. 
\end{equation}
%
Here $\theta_\pm(\omega)$ are the phase shifts of the 
complex reflectivity under $\vec{B}_\pm$ fields, which 
can be obtained from the $R_\pm(\omega)$ spectra using the 
Kramers-Kronig (K-K) relations.\cite{dressel}   
A superconducting magnet was used to apply the field, and 
a Fourier-transform interferometer was used to record the 
reflectivity spectra.    
The magneto-optical spectra were taken at photon energies 
below 1.1~eV, to which the zero-field spectra up to 
35~eV\cite{okamura} were smoothly connected.    

\section{Results and Discussion}
Figure 1 shows the $R_{avg}(\omega)$ and the MCD spectra 
of Tl$_2$Mn$_2$O$_7$ at 40~K for $B$=2, 4, and 6~T.   
$R_{avg}(\omega)$ at 40~K shows only minor dependence on 
$B$,\cite{okamura} and the $R_{avg}(\omega)$ spectra at 
0 and 6~T in Fig.~1 are almost unchanged.    
The sharp minimum in $R_{avg}(\omega)$ near 0.17~eV is the plasma edge, 
below which the reflectivity is high due to the Drude response 
of free carriers.     The sharp structures in $R_{avg}(\omega)$ 
below 0.08~eV are due to optical phonons.    
As shown in the bottom graph of Fig.~1, a clear MCD is observed 
around the plasma edge, which becomes stronger with increasing $B$.     
The sharp MCD peaks around 0.05~eV probably result 
from the plasmon-LO phonon coupling,\cite{lax1} 
which will not be discussed in this work.    
The Kerr rotation $\theta_K(\omega)$ at 6~T is also shown 
in Fig.~1.    It is seen that the maximum Kerr rotation 
of Tl$_2$Mn$_2$O$_7$ is about 1 degree at $B$=6~T.

The temperature dependence of the observed MCD at $B$=6~T 
is shown in Fig.~2, together with the $R_{avg}(\omega)$ spectra at 0 
and 6~T.    
The field-induced variations of $R_{avg}(\omega)$ at 125~K, 
near $T_C$, are very large.   
This is due to the increase of free carriers 
caused by the band polarization induced by the applied magnetic 
field,\cite{okamura} which is closely related with the CMR.   
Away from $T_C$, however, the field-induced changes in 
$R_{avg}(\omega)$ are very small.   
The observed MCD becomes stronger with decreasing $T$.   
In addition, the position of the MCD peak closely follows 
that of the plasma edge, as the latter undergoes blue 
shifts with decreasing $T$.    
This result strongly suggests that {\it the observed MCD is 
closely related with the free carriers} in Tl$_2$Mn$_2$O$_7$.     
Note that the field-induced changes of $R_{avg}(\omega)$ are 
largest near $T_C$, but the MCD becomes largest at lower 
temperatures.

Figure~3 plots the magnitude (peak height) of the MCD 
as functions of magnetic field and temperature, with the 
measured magnetization ($M$) of the same sample.     
Note that $M$ is induced even well above $T_C$ under applied 
fields of 4 and 6~T.    
Figure~3 clearly shows that the variation of MCD does not 
closely follow that of $M$.   
This is most clearly seen at 40~K in Fig.3(a), 
where $M$ is almost saturated above 0.5~T although 
the MCD grows with increasing $B$ even above 4~T.     
In addition, Fig.~3(b) shows that the MCD appears only below 160~K, 
which is coincident with a rapid increase of $M$ with cooling.  
This result probably indicates that a well-defined 
conduction band can be established only below 160~K, 
although an induced $M$ exists even at higher temperatures.

An MCD due to free carriers may arise through several mechanisms.   
In a ferromagnetic metal such as Gd, free carriers may lead to 
an MCD due to their skew scatterings caused by the internal $M$ 
and a spin-orbit coupling.\cite{Gd}     
Also for paramagnetic metals, a large MCD at $\omega_p$ has been 
attributed to an exchange-enhanced splitting of the plasma 
edge.\cite{schoenes}     In these cases, the MCD should be 
proportional to the spin polarization of the carriers.     
This is in contrast to the present results, 
where the observed MCD is not scaled with the internal $M$.   
Another possibility is a plasma edge-enhancement of 
MCD,\cite{feil,schoenes2,feil2} where a strong MCD is 
caused when a steep plasma edge is present and well 
separated in energy from the interband transitions.    
Although the $R(\omega)$ spectra of Tl$_2$Mn$_2$O$_7$ below 
$T_C$ may satisfy 
both of these conditions,\cite{okamura} this mechanism 
does not seem consistent with the $B$-dependence of the 
MCD for Tl$_2$Mn$_2$O$_7$.   
Interaction of the plasma resonance with interband 
transitions involving magnetic states\cite{degiorgi} 
is also unlikely, since $\omega_p$ is far apart from the 
interband transitions, located above 2~eV.\cite{okamura}     
We show below that the classical magnetoplasma (MP) 
resonance,\cite{lax1,lax2} which is a coupled Drude-cyclotron 
response of free carriers under $B$ field, can basically 
account for the observed MCD.     
In this model an MCD appears at the plasma edge 
due to its splitting into $\omega_p \pm \omega_c$, 
where $\omega_c$ is the cyclotron energy.

We take the $z$ axis normal to the sample surface, and the polarization 
vectors lie in the $xy$ plane.    Then in the classical MP model 
for isotropic band electrons, the components of the 
complex dielectric tensor are given as\cite{lax1,lax2} 
	\begin{eqnarray}
	\hat{\epsilon}_{xx}& =& 1 - \frac{\omega_p^2}{\omega} 
	\frac{(\omega+i \gamma)}{(\omega+i\gamma)^2-\omega_c^2} 
	\\
	\hat{\epsilon}_{xy}& =& -\hat{\epsilon}_{yx}=\frac{\omega_p^2}{\omega} 	\frac{i \omega_c}{(\omega+i\gamma)^2-\omega_c^2}, 
	\label{tensor}
	\end{eqnarray}
where $\omega_p$ and $\omega_c$ are the plasma and cyclotron 
energies, respectively, and $\gamma$ is the damping.    
The complex refractive indeces for $\pm$ field directions, 
$\hat{N}_\pm$, are the two eigen values of the complex 
refractive index tensor.\cite{MOKE}  They satisfy 
$\hat{N}_\pm^2 = \hat{\epsilon}_{xx} \pm i \hat{\epsilon}_{xy}$.   
By definition, they also satisfy 
$\hat{N}_\pm^2 = \hat{\epsilon}_\pm = \epsilon_{1\pm} + i \epsilon_{2\pm}$, 
where $\hat{\epsilon}_\pm$ are the complex dielectric 
functions of the MP for the $\pm$ field directions, and 
$\epsilon_{1\pm}$ and $\epsilon_{2\pm}$ are their 
real and imaginary parts, respectively.    Hence we have 
	\begin{equation}
	\hat{\epsilon}_{xx} \pm i \hat{\epsilon}_{xy} = 
	\epsilon_{1\pm} + i \epsilon_{2\pm},  
	\label{dielectric}
	\end{equation}
Substituting (3) and (4) into (5), we obtain 
	\begin{eqnarray}
	\epsilon_{1\pm} & = & 1-\frac{\omega_p^2}{\omega^2+\gamma^2}
	\mp \frac{\omega_p^2 \omega_c(\omega^2-\gamma^2)}{\omega(\omega^2+\gamma^2)^2}
	\\
	\epsilon_{2\pm} & = & \frac{\gamma \omega_p^2}{\omega(\omega^2+\gamma^2)}
	\pm \frac{2\gamma \omega_p^2 \omega_c}{(\omega^2+\gamma^2)^2}, 
	\label{dielectric2}
	\end{eqnarray}
where we have used $(\omega_c / \omega_p) \ll 1$, valid for the 
present case.  
The total dielectric function is expressed as\cite{dressel} 
	\begin{equation}
	\hat{\epsilon}_\pm^t = \hat{\epsilon}_{\pm} + 
	\hat{\epsilon}^{ib} + \hat{\epsilon}^{ph},
	\label{total}
	\end{equation}
where $\hat{\epsilon}^{ib}$ arises from the higher-energy interband 
transitions, which partially screens the plasma oscillations, 
and $\hat{\epsilon}^{ph}$ arises from the optical phonons.  
$R_\pm(\omega)$ are given as\cite{dressel} 
	\begin{equation}
	R_\pm(\omega) = \frac{(n_\pm-1)^2+k_\pm^2}{(n_\pm+1)^2 + k_\pm^2},
	\label{R}
	\end{equation}
where $n_\pm$ and $k_\pm$ are real and imaginary refractive 
indices, respectively, obtained by solving 
$\epsilon_{1\pm}^t = n_\pm^2 - k_\pm^2$ and 
$\epsilon_{2\pm}^t = 2 n_\pm k_\pm$.        
Finally, the MCD in this model is obtained by 
substituting (9) into (1).

We simulate the observed MCD spectra using the above model 
as follows.      First, we obtained the Drude parameters 
$\gamma$ and $\omega_p$ by fitting the observed $R_{avg}(\omega)$ 
based on the regular Drude reflectivity, i.e., that obtained 
by setting $\omega_c$=0 in (6)-(9).     
In doing so, we also used the classical Lorentz oscillator 
model\cite{dressel} for the phonon part $\epsilon^{ph}(\omega)$, 
and the actual $\epsilon^{ib}(\omega)$ obtained from the 
measured $R_{avg}(\omega)$ through the K-K relations.\cite{footnote}   
The observed $R_{avg}(\omega)$ spectra could be fitted well by this 
procedure, as shown in Fig.~4 (left axis).     
Then, we substituted the obtained $\gamma$ and $\omega_p$ into 
(6)-(9), and adjusted $\omega_c$ as a parameter so that 
the MCD calculated through (1) reproduces the observed MCD 
spectrum.      The fitted value of $\omega_c$ was used 
to calculate the effective mass $m^\ast$, using the 
relation $\hbar \omega_c = e B_i/m^\ast c$.    
Here, we have assumed that the internal field $B_i$ 
acting on the electrons is equal 
to the applied field.\cite{footnote-internal}   
Figure 4 shows the simulated reflectivity and MCD spectra 
at 6~T, with the used parameters shown in the 
caption, and Fig.~5 plots the obtained 
$m^\ast$ values at different values of $B$ and $T$.       
%
%
The simulated spectra in Fig.~4 have well reproduced 
the observed MCD spectra in Fig.~2, with the obtained 
$m^\ast / m_0$ values of $\sim$ 0.8 at 4~T, and 
$\sim$ 0.6 at 6~T.     
These values are very reasonable, since they are close to those 
predicted by the band calculations for ferromagnetic 
Tl$_2$Mn$_2$O$_7$,\cite{singh,mishra,shimakawa3} and also to 
those estimated previously from the effective carrier 
density.\cite{okamura}     
Since the parameters $\gamma$ and $\omega_p$ have been 
chosen to reproduce $R_{avg}(\omega)$, the only free parameter 
in simulating the MCD is $m^\ast$.  
Considering this fact and the simpleness of the model, 
the agreement between the data and the 
simulation is remarkable.   The decrease in $m^\ast$ from 
$B$= 4~T to 6~T is likely to result from changes in the 
band structure caused by the increase of $M$.    
In contrast, $m^\ast$ does not show significant 
temperature dependence both at 4 and 6~T.

Regarding the field dependence of the MCD, there seems to 
be a threshold for the onset of MCD, as seen in Fig.~3(a): 
the MCD grows with $B$ for 4 and 6~T, but 
is almost absent at 2~T.    
This is unexpected from the simple MP model, which predicts 
that the MCD at 2~T is about half that at 4~T.   
(Note that a well-defined conduction band exists for 
$T \ll T_C$ even at zero field.)    
The reason for this threshold behavior is unclear at the 
present, and it is likely to result from complications not 
included in the simple M-P model.   For example, the electrons 
experience scattering due to disorder such as defects 
and impurities, in addition to electron-phonon interaction.   
The degree of disorder-related scattering may vary with 
the field strength, depending on the relative magnitude 
of the magnetic length (cyclotron radius) to the spatial 
size of the disorder potential.      Such scattering may 
make the contribution of cyclotron motion to the MP 
resonance observed only above certain field strength.

Our previous work\cite{okamura} has shown that the 
variations of $R(\omega)$ 
and $\sigma(\omega)$ as functions of $B$ and $T$ closely follow 
the associated variations of the internal magnetization $M$.    
In fact, the effective carrier density evaluated from 
the measured spectra scaled with $M^2$ over wide ranges 
of $T$ and $B$.     In addition, the band calculations have 
predicted that the conduction electrons in Tl$_2$Mn$_2$O$_7$ are 
completely spin polarized.    In view of these previous results, 
the absence of an $M$-scaled MCD for Tl$_2$Mn$_2$O$_7$ is a rather 
unexpected result.       
Since the angular momentum carried by a circularly 
polarized photon cannot directly couple with the electron 
spin and hence with $M$, the carriers must have a sufficient 
spin-orbit ($s$-$o$) coupling to lead to an MCD.      
Therefore the present result strongly suggests a weak 
$s$-$o$ coupling for the conduction electrons in Tl$_2$Mn$_2$O$_7$.    
A weak $s$-$o$ coupling of carries in Tl$_2$Mn$_2$O$_7$ has also 
been indicated by the observation of a very weak anomalous Hall 
effect even well below $T_C$.\cite{imai}    
According to the band calculations, the density of states 
for the conduction band below the Fermi level in ferromagnetic 
Tl$_2$Mn$_2$O$_7$ are mainly derived from the Tl 6$s$, O 2$p$, 
and Mn 3$d$ states about 1:1:1 ratio.      Among these, 
the orbital angular momentum for 3$d$ electron is usually 
quenched, so that the O 2$p$ component may be the only 
source of $s$-$o$ coupling for the conduction electrons.    
This may account for the absence of an $M$-dependent MCD for 
Tl$_2$Mn$_2$O$_7$.      For a more quantitative account of 
the observed MCD, however, it is probably necessary 
to take into account the actual band structures, as well as 
their dependence on the magnetization.

It is interesting to compare the present results of Tl$_2$Mn$_2$O$_7$ 
with those of EuB$_6$,\cite{EuB6-PRL,EuB6-PRB,EuB6-EPJ} another 
ferromagnet with $T_C$=16~K which shows large magneto-optical effects 
similar to those for Tl$_2$Mn$_2$O$_7$.             
Near $T_C$, EuB$_6$ shows very large shifts of the plasma edge 
in $R(\omega)$ with $T$ and $B$.\cite{EuB6-PRL,EuB6-PRB}     
The Drude weight (effective carrier density) scales with $M^2$ 
over wide ranges of $T$ and $B$.\cite{EuB6-PRB}   
A pronounced MCD appears at the plasma edge of $R(\omega)$, 
with a Kerr rotation angle as large as 8 degrees.\cite{EuB6-EPJ}    
Although these results are qualitatively similar to those found 
for Tl$_2$Mn$_2$O$_7$, 
there is one marked difference concerning the plasma edge MCD: 
In addition to the much larger magnitude, 
the MCD for EuB$_6$ follows the variation of $M$ more closely 
than that for Tl$_2$Mn$_2$O$_7$.\cite{EuB6-EPJ}     
This is because the plasma edge MCD for EuB$_6$ 
is due to the spin polarization of the localized Eu 4$f$ 
electrons under $B$ fields, aided by the coupling between 
the Drude dynamics and the nearby interband transitions 
involving Eu 4$f$ states.\cite{EuB6-EPJ}     
This result is reasonable, since the Eu 4$f$ electrons are 
directly responsible for the $M$.   
In contrast, in the case of Tl$_2$Mn$_2$O$_7$ the source of MCD is the 
free carriers themselves, which have only small $s$-$o$ 
coupling resulting in the absence of $M$-scaled MCD.

\section*{Conclusion}
We have studied the infrared MCD (magneto-optical Kerr effect) 
of the magnetoresistive pyrochlore Tl$_2$Mn$_2$O$_7$, 
using synchrotron radiation source.    
A pronounced MCD signal has been observed exactly at the 
plasma edge of the reflectivity.    The observed MCD grows 
with the applied external magnetic field, and it is not 
scaled with the internal magnetization.     
The MCD has been successfully analyzed in terms of the 
classical magnetoplasma resonance model.     The absence 
of strongly $M$-dependent MCD in Tl$_2$Mn$_2$O$_7$ indicates 
a weak spin-orbit coupling of the conduction electrons 
in Tl$_2$Mn$_2$O$_7$.

\section*{Acknowledgements} 
This work was performed as a joint studies program of the 
Institute for Molecular Science (2001).    
This work was partly supported by Grants-In-Aid from the MEXT.

\newpage

\begin{figure}[t]
\begin{center}
\end{center}
\caption{
Top graph: the reflectivity ($R$) spectrum of 
Tl$_2$Mn$_2$O$_7$ at $T$=40~K and $B$= 0 and 6~T.  
Bottom panel: the MCD spectra, defined by (1), and 
the Kerr rotation $\theta_K$ measured at $T$=40~K 
for $B$=2, 4, and 6~T.  
}
\end{figure}

\begin{figure}[t]
\begin{center}
\end{center}
\caption{
Unpolarized reflectivity spectra at magnetic fields 
0~T (dotted curves) and 6~T (solid curves), and the MCD 
at 6~T (squares) of Tl$_2$Mn$_2$O$_7$ measured at four 
temperatures.    MCD signals below 0.06~eV have been 
omitted for clarity.   
}
\end{figure}

\begin{figure}[t]
\begin{center}
\end{center}
\caption{
The magnitude of measured MCD (circles, squares, and 
triangles) plotted on the right axis as functions of 
(a) magnetic field and (b) temperature.    
The magnetization ($M$) of the sample is also plotted on 
the left axis 
(solid, dotted, and dashed-dotted curves).    
}
\end{figure}

\begin{figure}[t]
\begin{center}
\end{center}
\caption{
Simulated reflectivity and MCD spectra at 6~T based on 
the MP model, as described in the text.    
The Drude parameters used are ($\omega_p$, $\gamma$)=
(0.24~eV, 52~meV) for 140~K, (0.325~eV, 39~meV) for 125~K, 
(0.38~eV, 30~meV) for 100~K, and (0.445~eV, 30~meV) for 
40~K.   See also the note.\cite{footnote2}
}
\end{figure}

\begin{figure}[t]
\begin{center}
\end{center}
\caption{
Effective masses in units of the rest electron mass 
($m^\ast / m_0$) obtained from the MCD data at magnetic 
fields $B$=4 and 6~T as a function of temperature.   
}
\end{figure}


\begin{thebibliography}{99}
\bibitem{footnote-mcd}  Actually, the incident light used 
was not completely circularly polarized, but elliptically 
polarized.    However, the degree of circular polarization 
was estimated to be greater than 90~\%.\cite{kimura-1}      
Hence the uncertainty in the measured MCD caused by the 
incomplete circular polarization is less than 10~\%.
\bibitem{footnote} 
In the measured spectra of Tl$_2$Mn$_2$O$_7$, the 
interband transition peaks were well separated in energy 
from the Drude-Lorentz components.\cite{okamura}    
Hence $\epsilon^{ib}(\omega)$ could be easily isolated from 
the measured total $\epsilon^t(\omega)$.    
The factor 1 in (5) was dropped since it should be included in 
$\epsilon^{ib}(\omega)$.\cite{dressel}
\bibitem{footnote-internal}   
In the presence of a  magnetiztion $\vec{M}$, the internal 
field $\vec{B}_{i}$ may differ from the applied field $\vec{B}_0$, 
since $\vec{M}$ produces an additional field $\vec{B}_m$, 
where $\vec{B}_{i}=\vec{B}_0 + \vec{B}_m$.    
The detail of $\vec{B}_m$ depends on the sample shape and 
$\vec{M}$.   
%
If there is a strong $\vec{B}_m$ in the Tl$_2$Mn$_2$O$_7$ 
sample, this should appear, well below $T_C$, as a very 
rapid increase of the (ordinary) Hall voltage for 
$B_0 <$ 0.5~T [see Fig. 3(a)], 
and it should scale with $M$ rather than with $B_0$.  
However, no such feature was observed in the experiments\cite{imai} 
for a sample having similar shape as that used in 
the present work.  
%
In addition, for a wide, plate-shaped sample with $\vec{M}$ 
and $\vec{B}_0$ perpendicular to the sample surface, 
it is well known that $\vec{B}_{i}=\vec{B}_0$.     
Our sample was a flat plate of $\sim 5 \times 5 \times 1$~mm$^3$.   
%
From these facts, we believe that $\vec{B}_{i} \simeq \vec{B}_0$ 
in our experiments.
\bibitem{footnote2}   
The $\omega_p$ values obtained by 
fitting are much larger than the observed plasma edge 
positions in $R(\omega)$, since the fitted values are ``bare'' 
plasma energies while those observed in $R(\omega)$ have been 
reduced due to the screening by $\epsilon^{ib}$.
\end{thebibliography}
\end{document}